\documentclass[11pt,eqs]{article}
\usepackage{latexsym,amsmath,amstext,amssymb}

\def\cleq{\setcounter{equation}{0}}

\title{
T-duality in the weakly curved background
\thanks{Work supported in part by
the Serbian Ministry of Education, Science and Technological Development, under contract No. 171031.}}
\author{Lj. Davidovi\'c \thanks{e-mail: ljubica@ipb.ac.rs} and B. Sazdovi\'c
\thanks{e-mail: sazdovic@ipb.ac.rs}\\
{\it Institute of Physics,}\\
{\it University of Belgrade,}\\
{\it 11001 Belgrade, P.O.Box 57, Serbia}}

\begin{document}
\maketitle

\begin{abstract}
We consider the closed string propagating
in the weakly curved background
which consists of constant metric and
Kalb-Ramond field with infinitesimally small
coordinate dependent part.
We propose the procedure for constructing the T-dual theory,
performing T-duality transformations along coordinates
on which the Kalb-Ramond field depends.
The obtained theory is defined in the non-geometric double
space, described by the
Lagrange multiplier $y_\mu$
and its $T$-dual in the flat space $\tilde{y}_\mu$.
We apply the proposed T-duality procedure to the T-dual theory
and obtain the initial one.
We discuss the standard relations between T-dual theories
that
the equations of motion and momenta modes of one theory
are the Bianchi identities and the winding modes of the other.
\end{abstract}
%%%%%%%%%%%%%%%%%%%%%%%%%%%%%%%%%%%%%%%%%%%%%%%%%%%%%%%%%%%%%%%%%%%
%%%%%%%%%%%%%%%%%%%%%%%%%%%%%%%%%%%%%%%%%%%%%%%%%%%%%%%%%%%%%%%%%%%
%%%%%%%%%%%%%%%%%%%%%%%%%%%%%%%%%%%%%%%%%%%%%%%%%%%%%%%%%%%%%%%%%%%%%%%
\section{Introduction}
\cleq

In string theory,
the duality symmetry was for the first time described in
the context of toroidal compactification in \cite{BKS}
(thoroughly explained in \cite{AAL}).
If only one dimension is compactified on the radius $R$,
then under the following transformation
\begin{equation}\label{eq:rR}
R\rightarrow \frac{\alpha^\prime}{R},\quad
\Phi\rightarrow \Phi-\log(\frac{R}{\sqrt{\alpha^\prime}}),
\end{equation}
where {$\alpha^\prime$ is the Regge slope parameter},
the physical features of the
interacting theory remain the same.
This kind of symmetry can be generalized to the arbitrary
toroidal compactification \cite{NSW},
and extended to the non-flat conformal backgrounds \cite{B}.
In the case of the open string,
there exists the relation between T-dual background fields
and the coordinate non-commutativity parameters \cite{SN},
as well as the relation between fermionic T-dual fields
and the parameters of the momenta non-commutativity \cite{SN1}.

In Buscher's construction of T-dual theory \cite{B,RV},
one starts with the manifold containing
the metric $G_{\mu\nu}$, the antisymmetric field $B_{\mu\nu}$
and the dilaton field $\Phi$.
It is required that the metric admits at least one continuous
abelian isometry which leaves the action for the $\sigma$-model
invariant.
One can choose
the target space coordinates
$x^\mu=(x^{i}, x^{a})$,
such that the isometry acts by the translation of the periodic
coordinates $x^{a}$.
The
T-duality along these directions changes
$x^{a}$ independent
background fields $G,\,B,\,\Phi$
into the corresponding T-dual fields $\tilde{G},\,\tilde{B},\,\tilde{\Phi}$.
In this way one connects different geometries
and two seemingly different $\sigma$-models.
This method was originally obtained in non covariant way
(because of the choice of coordinates),
but it was soon slightly modified,
which lead to the covariant construction \cite{AABL}.

In the covariant construction,
the isometry is gauged by
introducing the gauge fields $v_{\alpha}^{\mu}$.
In order to preserve the physical meaning of the original theory,
one requires that the new fields $v^\mu_\alpha$
do not carry the additional degrees of freedom.
This means that these fields
are pure gauge
with vanishing field strength
\begin{equation}
F^{\mu}_{\alpha\beta}=\partial_{\alpha}v^{\mu}_{\beta}
-\partial_{\beta}v^{\mu}_{\alpha}.
\end{equation}
This requirement
is included in the theory by adding the term
$y_{\mu}F^{\mu}_{01}$ into the Lagrangian, with $y_{\mu}$
being the Lagrange multiplier.
This guarantees that
at the classical level the dual theory will be
equivalent to the original one.
Integrating over the Lagrange multipliers $y_{\mu}$,
one
simply recovers the original theory.
The integration over the gauge fields $v_{\alpha}^{\mu}$,
produces the $T$-dual theory.
The non-abelian extension to T-duality has been considered in Refs. \cite{RJ,RD,RP,RS}.

In the construction above the background fields were
constant along $x^{a}$ directions.
In the present article, we consider
the weakly curved background.
We allow the background fields to depend on
the coordinates along which we perform duality transformations.
Note that the 
constant shift of coordinates remains the global symmetry
in this background.

In order to gauge the global isometry,
we introduce the gauge fields $v^\mu_\alpha$, as usual.
The replacement of the derivatives
$\partial_{\alpha}x^\mu$ with the
covariant ones $D_\alpha x^\mu$,
does not make the whole action invariant.
The obstacle is the background field $B_{\mu\nu}$ depending on $x^\mu$,
which is not locally gauge invariant.
So, in addition
we should covariantize $x^\mu$ as well.
At this point we will depart from the conventional approach.
We take
the invariant coordinate
as the line integral of the covariant derivatives
of the original coordinate
\begin{eqnarray}\label{eq:xi}
\Delta x^{\mu}_{inv}
&=&\int_{P}(d\xi^{+}D_{+}x^{\mu}
+d\xi^{-}D_{-}x^{\mu})
\nonumber\\
&=&
x^\mu-x^\mu(\xi_{0})+\Delta V^\mu[v_{+},v_{-}],
\end{eqnarray}
where $\Delta V^\mu$ is a line integral of the gauge fields $v^\mu_{\alpha}$.
As before, in order to obtain the theory physically equivalent
to the original one,
all degrees of freedom carried by the gauge fields $v^\mu_\alpha$
should be eliminated. Therefore,
we add Lagrange multiplier term
$y_{\mu}F^{\mu}_{01}$ into Lagrangian.
This allows us to consider $x^\mu_{inv}$ and $V^\mu$
as primitives up to a constant
of $D_\alpha x^\mu$ and $v^\mu_\alpha$ respectively.
Using the local gauge freedom we fix the gauge taking $x^\mu(\xi)=x^\mu(\xi_{0})$.

So, we succeed in generalizing the gauged action, which is
equal to
the original one
on the equations of motion for the Lagrange
multiplier.
The T-dual theory is obtained
on the equations of motion for the gauge fields $v^{\mu}_{\alpha}$.
These equations must be resolved iteratively, because 
\begin{enumerate}
 \item
the action is not bilinear in $v^\mu_{\alpha}$ as the background fields depend on
$V^\mu[v_{\alpha}]$
\item
$V^\mu$ is the line integral of $v^\mu_{\alpha}$.
\end{enumerate}

The fact that this background is characterized by the
infinitesimally small parameter enables one to solve the problem
and find the T-dual action. There are two essential differences in
the T-dual action in comparison to the flat background case. The
first one is that the target space of the T-dual theory in the
weakly curved background turns out to be the non-geometrical one
\cite{RP,RS,H}. This is doubled space with two coordinates, one of
them being the Lagrange multiplier as in the case of the flat
background. The second one is $T$-dual of the first in the flat space.
The second difference is the coordinate dependence of both dual
background fields as a consequence of the coordinate dependent
initial Kalb-Ramond field.

The theory defined above has one additional difficulty,
because
the invariant coordinate $\Delta x^\mu_{inv}$
and $\Delta V^\mu$ are defined as line integrals along arbitrary path $P$.
We will show that
the equation of motion for the Lagrangian multiplier $y_\mu$,
forces the field strength $F^\mu_{01}$ to vanish,
which guarantees that $\Delta x^\mu_{inv}$ and $\Delta V^\mu$ are independent on the choice of the path $P$.

Because T-duality leads to the equivalent theory,
we expected that the T-dual of the T-dual theory is the initial one.
The T-dual theory is
defined in doubled space but is still
globally invariant under
the shift of the T-dual coordinate $y_\mu$.
Gauging this symmetry, we show that the T-dual of the T-dual is indeed the original theory.

%%%%%%%%%%%%%%%%%%%%%%%%%%%%%%%%%%%%%%%%%%%%%%%%%%%%%%%%%%%%%%%%%%%%%%%%%%%%%%%%%%%%%%%%%%%%%%%%%%%%%
\section{
Bosonic string in the weakly curved background}
\cleq

Let us consider the action \cite{S}
\begin{equation}\label{eq:action0}
S= \kappa \int_{\Sigma} d^2\xi\Big[\frac{1}{2}\eta^{\alpha\beta}G_{\mu\nu}[x]
+\varepsilon^{\alpha\beta}B_{\mu\nu}[x]\Big]
\partial_{\alpha}x^{\mu}\partial_{\beta}x^{\nu},
\end{equation}
describing the propagation of the bosonic string in the non-trivial background,
defined by the space-time metric $G_{\mu\nu}$
and the Kalb-Ramond field $B_{\mu\nu}$.
The integration goes over two-dimensional world-sheet $\Sigma$
pa\-ra\-me\-tri\-zed by
$\xi^\alpha$ ($\xi^{0}=\tau,\ \xi^{1}=\sigma$),
where the intrinsic world-sheet metric $g_{\alpha\beta}$ is taken in the conformal gauge
$g_{\alpha\beta}=e^{2F}\eta_{\alpha\beta}$.
Here
$x^{\mu}(\xi),\ \mu=0,1,...,D-1$ are the coordinates of the
D-dimensional space-time,
$\kappa=\frac{1}{2\pi\alpha^\prime}$
and $\varepsilon^{01}=-1$.

Introducing the light-cone coordinates and their de\-ri\-va\-ti\-ves
\begin{equation}
\xi^{\pm}=\frac{1}{2}(\tau\pm\sigma),
\qquad
\partial_{\pm}=
\partial_{\tau}\pm\partial_{\sigma},
\end{equation}
and defining
\begin{eqnarray}
\Pi_{\pm\mu\nu}[x]=
B_{\mu\nu}[x]\pm\frac{1}{2}G_{\mu\nu}[x],
\end{eqnarray}
the action (\ref{eq:action0}) can
be written in the form
\begin{equation}\label{eq:action1}
S[x] = \kappa \int_{\Sigma} d^2\xi\
\partial_{+}x^{\mu}
\Pi_{+\mu\nu}[x]
\partial_{-}x^{\nu}.
\end{equation}

The consistency of the theory requires the conformal
invariance of the world-sheet on the quantum level.
This requirement results in  the space-time equations of motion for the
background fields. 
To the lowest order in slope parameter $\alpha^\prime$,
these equations have the form
\begin{eqnarray}\label{eq:beta}
&&
R_{\mu\nu}-\frac{1}{4}B_{\mu\rho\sigma}B_\nu^{\ \rho\sigma}+2D_\mu \partial_\nu\Phi=0,
\nonumber\\
&&
D_\rho B^\rho_{\ \mu\nu}-2\partial_\rho\Phi B^\rho_{\ \mu\nu}=0,
\nonumber\\
&&
4(\partial\Phi)^{2}-4D_\mu\partial^\mu\Phi
+\frac{1}{12}B_{\mu\nu\rho}B^{\mu\nu\rho}-R
=0\,,
\end{eqnarray}
where
$B_{\mu\nu\rho}=\partial_\mu B_{\nu\rho}
+\partial_\nu B_{\rho\mu}+\partial_\rho B_{\mu\nu}$
is the field strength of the field $B_{\mu \nu}$, and
$R_{\mu \nu}$ and $D_\mu$ are the Ricci tensor and the
covariant derivative with respect to the space-time metric.
$\Phi$ is dilaton field and $D$ is dimension of the space-time.
We consider
one of the simplest coordinate dependent solutions
of (\ref{eq:beta}).
This is the weakly curved background,
defined by the following expressions
\begin{eqnarray}\label{eq:wcb}
&&G_{\mu\nu}=const,
\nonumber\\
&&
B_{\mu\nu}[x]=b_{\mu\nu}+
\frac{1}{3}B_{\mu\nu\rho}x^\rho
\equiv 
b_{\mu\nu}+
h_{\mu\nu}[x],
\nonumber\\
&&\Phi=const.
\end{eqnarray}
This background is the solution of 
the space-time equations of motion
if the constant
$B_{\mu\nu\rho}$ is taken to be infinitesimally small
and all the calculations are done in the first order
in $B_{\mu\nu\rho}$.

The weakly curved background was considered in Refs.\cite{DS},
where the influence of the boundary conditions on the non-commutativity
of the open bosonic string has been investigated.
The same approximation (not refereed to as weakly curved background),
was considered in \cite{ALLP}, where the closed string non-com\-mu\-ta\-ti\-vity
was investigated. 
In the present paper, we will investigate the closed bosonic string
moving in the weakly curved background,
with a goal to find the generalization of the
Buscher's construction of the T-dual theory.

%%%%%%%%%%%%%%%%%%%%%%%%%%%%%%%%%%%%%%%%%%%%%%%%%%%%%%%%%%%%%%%%%%%%%%%%%%%%%%%%%%%%
\section{
Generalized Buscher's construction
}\label{sec:construction}
\cleq

In the standard Buscher's construction of T-dual theory,
the premise
is that the target space has isometries.
It is possible to choose adopted coordinates $x^\mu = (x^i , x^a)$,
so that the isometries act as translations of $x^a$
components
and as the background fields are taken to be $x^a$-independent,
the action is invariant under the global shift symmetry.
The weakly curved background preserves this symmetry,
despite of $x^a$-dependence of the background fields.
As this is not obvious, let us first demonstrate that the constant shift
\begin{equation}
\delta x^\mu=\lambda^\mu=const,
\end{equation}
leaves the action (\ref{eq:action1}) for the closed string, invariant.
For simplicity we assume that all the coordinates are compact.

As $B_{\mu\nu}$ is linear in coordinate,
one has
\begin{eqnarray}\label{eq:var}
\delta S&=&
\frac{\kappa}{3}B_{\mu\nu\rho}\lambda^\rho
\int d^{2}\xi
\partial_{+}x^\mu \partial_{-}x^\nu
\nonumber\\
&=&\frac{\kappa}{3}B_{\mu\nu\rho}\lambda^\rho
\epsilon^{\alpha\beta}
\int d^{2}\xi \partial_{\alpha}x^\mu \partial_{\beta}x^\nu.
\end{eqnarray}
This is proportional to the total divergence
\begin{equation}\label{eq:vars}
\delta S=
\frac{\kappa}{3}B_{\mu\nu\rho}\lambda^\rho
\epsilon^{\alpha\beta}
\int d^{2}\xi\partial_{\alpha}(x^\mu\partial_{\beta}x^\nu)
=0,
\end{equation}
which vanishes in the case of the closed string
and the topologically trivial mapping of the world-sheet into the space-time.

%%%%%%%%%%%%%%%%%%%%%%%%%%%%%%%%%%%%%%%%%%%%%%%%%%%%%%%%%%%%%%%%%%%%%%%%%%%%%%%%%%%%%%%%
\subsection{Gauging shift symmetry}

In the present paper the procedure for gauging the global shift
symmetries is different from the conventional one
\cite{B,RP,RS,RC}. The coordinate dependence of the Kalb-Ramond
field,
separates us from the conventional approach.

To localize the global symmetry, we introduce the independent world-sheet
gauge fields $v^{\mu}_{\alpha}$ and
substitute the ordinary derivatives with the covariant ones
\begin{equation}
\partial_{\alpha}x^\mu\rightarrow
D_{\alpha}x^\mu=\partial_\alpha x^\mu
+v^{\mu}_{\alpha}.
\end{equation}
We want the covariant derivatives to be gauge invariant,
so we impose the transformation law for the gauge fields
\begin{equation}\label{eq:gft}
\delta v^{\mu}_{\alpha}=-\partial_\alpha \lambda^\mu,\quad
(\lambda^\mu=\lambda^\mu(\tau,\sigma)).
\end{equation}
This replacement is, however, not
sufficient to make the action
locally invariant because
the background field $B_{\mu\nu}$ in
the weakly curved background,
depends on the coordinate $x^\mu$
which is not gauge invariant.
So, there is one more important step to be done.
We should replace the coordinate
$x^\mu$,
with some extension for it,
where only already introduced gauge fields
$v^{\mu}_{\alpha}$ will appear.
Let us define the invariant coordinate by
\begin{eqnarray}
\Delta x^{\mu}_{inv}&\equiv&
\int_{P}d\xi^\alpha\,D_\alpha x^\mu
=\int_{P}(d\xi^{+}D_{+}x^{\mu}
+d\xi^{-}D_{-}x^{\nu})
\nonumber\\
&=&
x^\mu-x^\mu(\xi_{0})
+\Delta V^\mu,
\end{eqnarray}
where
\begin{equation}\label{eq:vdef}
\Delta V^\mu
\equiv
\int_{P}d\xi^\alpha v^{\mu}_{\alpha}
=\int_{P}(d\xi^{+} v^{\mu}_{+}
+d\xi^{-} v^{\mu}_{-}).
\end{equation}
The  line integral is
taken along the path $P$,
from the initial point $\xi^{\alpha}_{0}(\tau_{0},\sigma_{0})$
to the final one $\xi^{\alpha}(\tau,\sigma)$.

We require the T-dual theory to be
equivalent to the initial one.
So, we do not want to introduce new degrees of freedom,
originating from the gauge fields.
Therefore, we will require the corresponding field strength
\begin{equation}
F^{\mu}_{\alpha\beta}\equiv\partial_{\alpha}v^{\mu}_{\beta}
-\partial_{\beta}v^{\mu}_{\alpha},
\end{equation}
to vanish.
We can achieve this by introducing the
Lagrange multiplier $y_\mu$,
and the appropriate term in the Lagrangian
which will force $F^{\mu}_{+-}\equiv
\partial_{+}v^{\mu}_{-}
-\partial_{-}v^{\mu}_{+}=-2F^{\mu}_{01}$ to vanish.
So, the gauge invariant action is
\begin{eqnarray}\label{eq:ainv}
S_{inv}&=&\kappa\int d^{2}\xi\Big{[}
D_{+}x^\mu \Pi_{+\mu\nu}[\Delta x_{inv}] D_{-}x^\nu
+\frac{1}{2}(v^{\mu}_{+}\partial_{-}y_\mu
-v^{\mu}_{-}\partial_{+}y_\mu)
\Big{]},
\end{eqnarray}
where the last term is equal
$\frac{1}{2}y_\mu F^{\mu}_{+-}$ up to the total divergence.
Now, we can use the gauge freedom to fix the gauge.
It is easy to see that $x^\mu(\xi)=x^\mu(\xi_{0})$ is good gauge fixing.
So, the gauge fixed action equals
\begin{eqnarray}\label{eq:sfix}
S_{fix}[y,v_{\pm}]&=&\kappa\int d^{2}\xi\Big{[}
v_{+}^\mu \Pi_{+\mu\nu}[\Delta V] v_{-}^\nu
+\frac{1}{2}(v^{\mu}_{+}\partial_{-}y_\mu
-v^{\mu}_{-}\partial_{+}y_\mu)
\Big{]},
\end{eqnarray}
where $y_\mu$ and $v_{\pm}^\mu$ are independent variables and
$\Delta V^\mu$ is defined in (\ref{eq:vdef}).

Note that,
we can also define $x^\mu_{inv}$ and $V^\mu$ as the solutions of the equations 
$\partial_\alpha x^\mu_{inv}=D_\alpha x^\mu$ and $\partial_\alpha V^\mu=v^\mu_\alpha$.
So, $x^\mu_{inv}$ and $V^\mu$ 
are in fact primitives and due to the presence of the term $\frac{1}{2}y_\mu F^{\mu}_{+-}$
in the Lagrangian they
are well defined up to the constant.
%%%%%%%%%%%%%%%%%%%%%%%%%%%%%%%%%%%%%%%%%%%%%%%%%%%%%%%%%%%%%%%%%%%%%%%%%%%%%%%%%%%%%%%%%%%%%%
\section{From gauge fixed action to the original and T-dual action}
\cleq

From the gauge fixed action (\ref{eq:sfix}),
we can obtain equations of motion varying by
Lagrange multiplier $y_\mu$ and the gauge fields $v^\mu_\pm$.
On the equation of motion for the Lagrange multiplier we will
obtain the original action,
and on the equations of motion for the gauge fields
we will obtain the T-dual theory.

%%%%%%%%%%%%%%%%%%%%%%%%%%%%%%%%%%%%%%%%%%%%%%%%%%%%%%%%%%%%%%%%%%%%%%%%%%%%%%%%%%%%%%%
\subsection{Eliminating the Lagrange multiplier }\label{sec:ilm}

Let us show that on the equation of motion for the
Lagrange multiplier
\begin{equation}\label{eq:emy}
\partial_{+}v^{\mu}_{-}
-\partial_{-}v^{\mu}_{+}=0,
\end{equation}
the gauge fixed action (\ref{eq:sfix}) reduces to the
original action (\ref{eq:action1}).
This equation of motion enforces
the field strength of the gauge fields $F^{\mu}_{+-}$ to vanish,
and therefore makes the variable $\Delta V^\mu$ defined in (\ref{eq:vdef})
path independent. To prove this let us show that
$\Delta V^\mu$  is equal to zero for the closed path.
If $P$ is closed path, then using Stoke's theorem
the defining integral along $P$,
can be rewritten as the integral over the surface $S$
which spans the path $P=\partial S$,
of the field strength of the gauge fields
\begin{equation}
\oint_{P=\partial S}
d\xi^{\alpha} v_{\alpha}^{\mu}
=\int_{S}d^{2}\xi\
(\partial_{+}v^{\mu}_{-}
-\partial_{-}v^{\mu}_{+}
),
\end{equation}
which is obviously zero on (\ref{eq:emy}). 

The solution of (\ref{eq:emy})
\begin{equation}\label{eq:vsol}
v^{\mu}_{\pm}=\partial_{\pm}x^\mu,
\end{equation}
substituted into (\ref{eq:vdef})
gives
\begin{equation}\label{eq:vvelx}
\Delta V^\mu(\xi)=x^\mu(\xi)-x^\mu(\xi_{0}).
\end{equation}
Taking into account that the action does not depend on the constant shift of the coordinate,
we can omit $x^\mu(\xi_{0})$ and so the action (\ref{eq:sfix})
becomes
\begin{eqnarray}\label{eq:ay}
S_{fix}[v_{\pm}=\partial_{\pm}x]&=&
\kappa\int d^{2}\xi\
{\partial_{+}x}^{\mu}\Pi_{+\mu\nu}[x]{\partial_{-}x}^{\nu},
\end{eqnarray}
which is just the initial action (\ref{eq:action1}).
%%%%%%%%%%%%%%%%%%%%%%%%%%%%%%%%%%%%%%%%%%%%%%%
\subsection{Eliminating the gauge fields}\label{sec:gfe}

The T-dual action will be obtained
by integrating out the gauge fields from (\ref{eq:sfix}).
The equations of motion with respect to the
gauge fields $v^{\mu}_{\pm}$ are
\begin{equation}\label{eq:emv}
\Pi_{\mp\mu\nu}[\Delta V]v_{\pm}^{\nu}
+\frac{1}{2}\partial_{\pm}y_\mu=
\mp\beta^{\mp}_\mu[V],
\end{equation}
with the terms $\beta^{\mp}_\mu[V]$ defined 
by
\begin{equation}\label{eq:betapm}
\beta^{\pm}_\mu[x]=
\mp\frac{1}{2}h_{\mu\nu}[x]\partial_\mp x^\nu.
\end{equation}

Notice that $\beta^{\mp}_\mu$
come from the variation
with respect to $\Delta V^\mu(\xi)$,
the argument of the background fields.
To show this,
let us find the variation with respect to $V^\mu$ (which depends on $v^\mu_\pm$)
\begin{eqnarray}\label{eq:svar}
\delta_{V}\, S_{fix}&=&
\kappa\int d^{2}\xi\,
 v^{\nu}_{+}\partial_\mu B_{\nu\rho}v^{\rho}_{-}
\,\delta V^\mu
\nonumber\\
&\equiv&\kappa\int d^{2}\xi\, \eta_\mu
\delta V^\mu,
\end{eqnarray}
where with the help of the relation
\begin{equation}\label{eq:vV}
\partial_\alpha V^{\mu}=
v^{\mu}_\alpha,
\end{equation}
we have
\begin{equation}\label{eq:eta}
\eta_\mu
=\partial_\mu B_{\nu\rho}\varepsilon^{\alpha\beta}\partial_{\alpha} V^{\nu}\partial_{\beta} V^{\rho}.
\end{equation}
So, we can write
\begin{equation}\label{eq:etabeta}
\eta_\mu=\partial_\alpha\beta^\alpha_\mu[V],
\quad
\beta^{\alpha}_{\mu}[x]\equiv
-\varepsilon^{\alpha\beta}
h_{\mu\nu}[x]\partial_{\beta}x^\nu,
\end{equation}
with $h_{\mu\nu}$ defined in (\ref{eq:wcb})
and consequently
\begin{eqnarray}
\delta_{V}\, S_{fix}&=&-\kappa\int d^{2}\xi \beta^\alpha_\mu[V] \delta v^\mu_\alpha
\nonumber\\
&=&-\kappa\int d^{2}\xi\Big{[}
\beta^{+}_\mu[V] \delta v^\mu_{+}+
\beta^{-}_\mu[V] \delta v^\mu_{-}\Big{]},
\end{eqnarray}
where $\beta^{\pm}_\mu[x]=\frac{1}{2}\big{(}
\beta^{0}_\mu[x]\pm\beta^{1}_\mu[x]
\big{)}$
is defined in (\ref{eq:betapm}).

Because $V^\mu$ is function of
$v^\mu_{+}$ and $v^\mu_{-}$,
there are two equations in (\ref{eq:emv})
with two unknown variables
$v^\mu_{+}$ and $v^\mu_{-}$.
We can rewrite (\ref{eq:emv}) in the form
\begin{eqnarray}\label{eq:vsol1}
v_{\pm}^{\mu}(y)&=&
-\kappa\,{\Theta}^{\mu\nu}_{\pm}[\Delta V(y)]
\Big{[}
\partial_{\pm}y_\nu
\pm2\beta^{\mp}_\nu[V(y)]
\Big{]},
\end{eqnarray}
where
\begin{eqnarray}\label{eq:velikateta}
{\Theta}^{\mu\nu}_{\pm}[\Delta  V]
&=&-\frac{2}{\kappa}
\Big(G^{-1}_{E}[\Delta  V]\Pi_{\pm}[\Delta  V]G^{-1}\Big)^{\mu\nu}
\nonumber\\
&=&\theta^{\mu\nu}[\Delta  V]\mp\frac{1}{\kappa}(G_{E}^{-1})^{\mu\nu}[\Delta V],
\end{eqnarray}
and
$G^{E}_{\mu\nu}\equiv[G-4BG^{-1}B]_{\mu\nu}$,
$\theta^{\mu\nu}\equiv
-\frac{2}{\kappa}
(G^{-1}_{E}BG^{-1})^{\mu\nu}$
are the open string background fields:
the effective metric and the non-commutativity parameter respectively.
Let us mention that the variables 
$G_{E}^{\mu\nu}$ and ${\theta}^{\mu\nu}$
correspond to the new fields $\tilde{g}$ and $\beta$ introduced by field redefinition
in \cite{TG}.
Tensors $\Pi_{\mp\mu\nu}$ and $\Theta^{\mu\nu}_{\pm}$
are connected by the relation
\begin{equation}\label{eq:tp}
\Theta^{\mu\nu}_{\pm}\Pi_{\mp\nu\rho}=
\frac{1}{2\kappa}\delta^{\mu}_{\rho}.
\end{equation}

We will solve eqs. (\ref{eq:vsol1}) iteratively.
Let us separate the variables into two parts,
similarly as in \cite{ALLP}
\begin{equation}\label{eq:nulajedan}
v_{\pm}^\mu=v_{\pm}^{(0)\mu}+v_{\pm}^{(1)\mu},
\quad
y_\mu=y_\mu^{(0)}+y_\mu^{(1)},
\end{equation}
where the index $(0)$ marks the finite part and index $(1)$ infinitesimal
part (proportional to $B_{\mu\nu\rho}$). 
In the zeroth order equations (\ref{eq:vsol1}) reduce to
\begin{equation}\label{eq:v0}
v_{\pm}^{(0)\mu}(y)=
-\kappa\,{\Theta}^{\mu\nu}_{0\pm}
\partial_{\pm}y_\nu^{(0)},
\end{equation}
where
\begin{eqnarray}\label{eq:tvnula}
{\Theta}^{\mu\nu}_{0\pm}
=-\frac{2}{\kappa}
(g^{-1}\Pi_{0\pm}G^{-1})^{\mu\nu}
=\theta^{\mu\nu}_{0}\mp\frac{1}{\kappa}(g^{-1})^{\mu\nu},
\end{eqnarray}
with $g_{\mu\nu}=G_{\mu\nu}-4b^{2}_{\mu\nu}$
and ${\theta}^{\mu\nu}_{0}
=-\frac{2}{\kappa}
(g^{-1}bG^{-1})^{\mu\nu}
$.
Tensors $\Pi_{0\mp\mu\nu}$ and $\Theta^{\mu\nu}_{0\pm}$
are connected by the relation 
\begin{equation}\label{eq:ptnula}
\Pi_{0\mp\mu\nu}\Theta^{\nu\rho}_{0\pm}=
\frac{1}{2\kappa}\delta_{\mu}^{\rho},
\end{equation}
which is the
reduction of (\ref{eq:tp}) to a constant background case.

In order to
explicitly express 
the background fields argument in zeroth order 
$V^{(0)\mu}$,
we introduce the
new (double) variable $\tilde{y}_\mu$ in the zeroth order
by
\begin{equation}\label{eq:tildey}
\Delta\tilde{y}^{(0)}_\mu=
\tilde{y}^{(0)}_\mu(\xi)-\tilde{y}^{(0)}_\mu(\xi_{0})
=\int_{P}d\tau y^{(0)\prime}_\mu+d\sigma\dot{y}^{(0)}_\mu,
\end{equation}
where the line integral is
independent of the choice of the path $P$
on the equation of motion.
The double variable satisfies
\begin{equation}
\dot{{\tilde{y}}}^{(0)}_\mu=
{y}_\mu^{(0)\prime},\quad
\tilde{y}_\mu^{(0)\prime}=\dot{y}_\mu^{(0)}.
\end{equation}
Using the last relation  and (\ref{eq:v0}) we obtain 
\begin{equation}\label{eq:argnula}
V^{(0)\mu}=
-\kappa\,{\theta}^{\mu\nu}_{0} y_\nu^{(0)}
+(g^{-1})^{\mu\nu}\tilde{y}_\nu^{(0)}.
\end{equation}

Comparing (\ref{eq:vsol}) with (\ref{eq:vsol1}) where the background fields argument is taken in the zeroth order (\ref{eq:argnula}),
we obtain the T-dual transformation law of the variables
\begin{equation}\label{eq:tl}
\partial_\pm x^\mu\cong
-\kappa\,{\Theta}^{\mu\nu}_{\pm}[\Delta V^{(0)}]
\partial_{\pm}y_\nu
\mp2\kappa {\Theta}^{\mu\nu}_{0\pm}\beta^{\mp}_\nu[V^{(0)}].
\end{equation}
In the flat background for $b_{\mu\nu}=0$ we have
$\tilde{y}_\mu\cong G_{\mu\nu} x^\nu$.
So, the double variable $\tilde{y}_\mu$ in this particular case turns to be related to $x^\mu$, the T-dual variable of $y_\mu$.

Substituting (\ref{eq:vsol1}) and (\ref{eq:argnula}) into the action (\ref{eq:sfix}),
we obtain T-dual action
\begin{equation}\label{eq:dualna}
^\star S[y]\equiv S_{fix}[y]=
\frac{\kappa^{2}}{2}
\int d^{2}\xi\
\partial_{+}y_\mu
\Theta_{-}^{\mu\nu}[\Delta V^{(0)}(y)]
\partial_{-}y_\nu,
\end{equation}
where we neglected the term $\beta^{-}_{\mu}\beta^{+}_{\nu}$
as the infinitesimal of the second order.

Comparing the initial action (\ref{eq:action1}) with
the T-dual one (\ref{eq:dualna}),
we see that they are equal under following transformations
\begin{equation}\label{eq:dual}
\partial_{\pm}x^\mu\rightarrow
\partial_{\pm}{y_\mu},\qquad
\Pi_{+\mu\nu}[x]\rightarrow
\frac{\kappa}{2}\Theta_{-}^{\mu\nu}[\Delta V^{(0)}
],
\end{equation}
which implies
\begin{eqnarray}\label{eq:transition}
&G_{\mu\nu}&\rightarrow\,
{^\star G}^{\mu\nu}=
(G^{-1}_{E})^{\mu\nu}[\Delta V^{(0)}
],
\nonumber\\
&B_{\mu\nu}[x]&\rightarrow\,
{^{\star}B}^{\mu\nu}=
\frac{\kappa}{2}\theta^{\mu\nu}[\Delta V^{(0)}
],
\end{eqnarray}
where
$(G^{-1}_{E})^{\mu\nu}$ and $\theta^{\mu\nu}$ are introduced in (\ref{eq:velikateta})
and
\begin{equation}
\Delta V^{(0)\mu}(y)
=-\kappa\theta_{0}^{\mu\nu}\Delta y^{(0)}_\nu+(g^{-1})^{\mu\nu}\Delta\tilde{y}^{(0)}_\nu.
\end{equation}
Let us underline
that in the initial theory the metric tensor
is constant and the Kalb-Ramond field is linear in coordinate $x^\mu$.
In the T-dual theory, both background fields depend on $\Delta V^\mu$,
which is the linear combination of $y_\mu$ and its dual ${\tilde{y}}_\mu$.
Note that the variable $V^\mu$
and consequently T-dual action is not defined on the geometrical space
(defined by the coordinate $y_\mu$)
but on the so called
doubled target space \cite{H} composed of both $y_\mu$ and $\tilde{y}_\mu$.
The similar procedure,
using the first order Lagrangian was
applied to the flat space-time, long time ago \cite{Duff}.
The result has the same form as (\ref{eq:transition}), but it is $V$-independent.

In the \ref{sec:dodatak},
the equation of motion for the
T-dual theory will be given explicitly.
It will be shown that this equation is equal to the equation of motion
of the gauge fixed action
after the elimination of the gauge fields as auxiliary fields on
their equation of motion.

%%%%%%%%%%%%%%%%%%%%%%%%%%%%%%%%%%%%%%%%%%%%%%%%%%%%%%%%%%%%%%%%%%%%%%%%%%%%%%%%%%%%%%%%%%%%%%%%

\section{The T-dual of the T-dual theory}
\cleq

Because the T-dual theory
(\ref{eq:dualna})
is by construction
physically equivalent to the initial one (\ref{eq:action1}),
we should expect that the T-dual of the T-dual
theory is just the initial theory.
To demonstrate this we should first find
the global symmetry of the T-dual action.
As can be seen from (\ref{eq:sfix}) the gauge fixed action, $S_{fix}$, is invariant under the global shift
\begin{equation}\label{eq:gsy}
\delta y_\mu=\lambda_\mu=const.
\end{equation}
As the T-dual theory is equivalent to the gauge fixed one, it must have the same
global symmetry.
One can check that this is indeed the symmetry of (\ref{eq:dualna}).
Note that the action is not invariant under the constant shift of the argument of $\Theta^{\mu\nu}_{-}$,
because, 
in contrast to the original action,
the metric of the T-dual theory $^\star G^{\mu\nu}$
is not constant.
So,
in comparison to the original action,
one can not omit the constant part of the argument $V^\mu(\xi_{0})$
as we did in the end of the subsection \ref{sec:ilm}.
But, the transformation (\ref{eq:gsy}) leaves the argument itself, $\Delta V^\mu=V^\mu(\xi)-V^\mu(\xi_{0})$, unchanged
and consequently the action (\ref{eq:dualna})
is invariant too.

%%%%%%%%%%%%%%%%%%%%%%%%%%%%%%%%%%%%%%%%%%%%%%%%%%%%%%%%%%%%%%%%%%%%%

\subsection{ Gauging the symmetry}

Let us localize this symmetry
and find the
corresponding locally invariant action.
We covariantize the derivatives introducing the
gauge fields $u_{\pm\mu}$
\begin{equation}
D_{\pm}y_\mu=\partial_{\pm}y_\mu+u_{\pm\mu}.
\end{equation}
Demanding
$
\delta D_{\pm}y_\mu=0,
$
we require that $u_{\pm\mu}$ transform as
\begin{equation}
\delta u_{\pm\mu}=-\partial_{\pm}\lambda_\mu(\tau,\sigma).
\end{equation}

The dual background fields argument $\Delta V^\mu$ is not locally invariant.
So, first we construct the invariant expressions for both variables $y_
\mu$ and $\tilde{y}_\mu$
\begin{eqnarray}
\Delta y^{inv}_\mu&\equiv&
\int_{P}d\xi^\alpha\,D_\alpha y_\mu
=\Delta y_\mu+\Delta U_\mu,
\nonumber\\
\Delta \tilde{y}^{inv}_\mu&\equiv&
\int_{P}d\xi^\alpha
\varepsilon^{\beta}_{\ \alpha}
 \,D_\beta y_\mu
=\Delta \tilde{y}_\mu
+\Delta \tilde{U}_\mu,
\end{eqnarray}
where
\begin{eqnarray}\label{eq:ytilde}
\Delta y_\mu&\equiv&
\int_{P} d\xi^\alpha \partial_\alpha y_\mu
=y_\mu(\xi)-y_\mu(\xi_{0}),
\nonumber\\
\Delta {\tilde{y}}_\mu&\equiv&
\int_{P} d\xi^\alpha \varepsilon^\beta_{\ \alpha}\partial_{\beta}y^\mu,
\end{eqnarray}
and
\begin{eqnarray}\label{eq:Udef}
\Delta U_\mu
\equiv
\int_{P}d\xi^\alpha u_{\alpha\mu},
\quad
\Delta \tilde{U}_\mu
\equiv
\int_{P}d\xi^\alpha \varepsilon^{\beta}_{\ \alpha} u_{\beta\mu}.
\end{eqnarray}

Now, it is easy to find the generalization of the background fields argument (\ref{eq:transition})
\begin{eqnarray}
\Delta V^\mu_{inv}&\equiv&
-\kappa\theta^{\mu\nu}_{0}\Delta y^{inv}_\nu
+(g^{-1})^{\mu\nu}\Delta\tilde{y}^{inv}_\nu
\nonumber\\
&=&\Delta V^\mu[y]+\Delta V^\mu[U],
\end{eqnarray}
which is invariant by construction,
and will be considered only in the zeroth order.

Finally, we can construct the dual invariant action
\begin{eqnarray}
^\star S_{inv}&=&
\frac{\kappa}{2}\int d^{2}\xi\Big{[}
\kappa D_{+}y_\mu\Theta^{\mu\nu}_{-}[\Delta V_{inv}]D_{-}y_\nu
+u_{+\mu}\partial_{-}z^\mu
-u_{-\mu}\partial_{+}z^\mu\Big{]},
\end{eqnarray}
where the second term makes the gauge fields $u_{\pm\mu}$ nonphysical.
The gauge fixing $y_{\mu}(\xi)=y_{\mu}(\xi_{0})$,
produces
$D_{\pm}y_{\mu}=u_{\pm\mu}$
and $\Delta V^\mu[y]=0$, so the action becomes
\begin{eqnarray}\label{eq:afix}
^\star S_{fix}[z,u_{\pm}]&=&
\frac{\kappa}{2}\int d^{2}\xi
\Big{[}\kappa
u_{+\mu}\Theta^{\mu\nu}_{-}\big{[}\Delta V[U]\big{]}u_{-\nu}
+
u_{+\mu}\partial_{-}z^\mu
-u_{-\mu}\partial_{+}z^\mu\Big{]}.
\end{eqnarray}
%%%%%%%%%%%%%%%%%%%%%%%%%%%%%%%%%%%%%%%%%%%%%%%%%%%%%%%%%%%%%%%%%%%%%%%%

\subsection{Eliminating the Lagrange multiplier}

The equation of motion
with respect to the Lagrange multiplier $z^\mu$
\begin{equation}
\partial_{+}u_{-\mu}
-\partial_{-}u_{+\mu}=0,
\end{equation}
has the solution
\begin{equation}\label{eq:usol}
u_{\pm\mu}=\partial_{\pm}y_\mu,
\end{equation}
which substituted to (\ref{eq:Udef}) gives
$\Delta U_\mu=\Delta y_\mu$ and therefore $\Delta V^\mu[U]=\Delta V^\mu[y]$.
So, the action (\ref{eq:afix}) on this solution becomes
\begin{equation}
^\star S_{fix}[u_{\pm}=\partial_{\pm}y]=
\frac{\kappa^{2}}{2}\int d^{2}\xi
\partial_{+}y_\mu
\Theta^{\mu\nu}_{-}\big{[}\Delta V[y]\big{]}
\partial_{-}y_\nu,
\end{equation}
and coincides with the T-dual action (\ref{eq:dualna}).

Let us stress that we can not omit $V(\xi_{0})$,
because,
as we have discussed at the beginning of the section,
the T-dual action is invariant under the constant shift in coordinate $y_\mu$,
but it is not invariant under the constant shift of the argument of the background fields.

%%%%%%%%%%%%%%%%%%%%%%%%%%%%%%%%%%%%%%%%%%%%%%%%%%%%%%%%%%%%%%%%%%%%%
\subsection{Eliminating the gauge fields}

Using the fact that
\begin{equation}
\Theta^{\mu\nu}_{\pm}(x)
=
\Theta^{\mu\nu}_{0\pm}
-2\kappa [\Theta_{0\pm}h(x)\Theta_{0\pm}]^{\mu\nu},
\end{equation}
we find that
the equations of motion for the gauge fields,
obtained by varying the action (\ref{eq:afix})
with respect to the gauge fields $u_{\pm\mu}$,
are
\begin{equation}
\partial_{\pm}z^\mu=-\kappa\Theta^{\mu\nu}_{\pm}[\Delta V(U)]
\Big{[}u_{\pm\nu}\pm 2\beta^{\mp}_{\nu}[V(U)]\Big{]}.
\end{equation}
Note that $\Theta^{\mu\nu}_{\pm}$ depends on $\Delta V^\mu(U)$ while $\beta_\nu$ depends on $V^\mu(U)$.
Using the relation
(\ref{eq:tp}),
we can extract $u_{\pm\mu}$
\begin{equation}\label{eq:upmz}
u_{\pm\mu}=-2\Pi_{\mp\mu\nu}[\Delta V(U)]\partial_{\pm}z^\nu
\mp 2\beta^{\mp}_{\mu}[V(U)].
\end{equation}
Similarly
as in Sect. \ref{sec:gfe},
we can sepparate the variables $u_{\pm\mu}$ into finite and infinitesimal part.
In the zeroth order (\ref{eq:upmz}) reduce to
\begin{equation}
u_{\pm\mu}^{(0)}=-2\Pi_{0\mp\mu\nu}\partial_{\pm}z^{(0)\nu}.
\end{equation}
So,
the zeroth order values of $U_\mu$ and $\tilde{U}_\mu$ are
\begin{eqnarray}
U_\mu^{(0)}&=&-2b_{\mu\nu}z^{(0)\nu}+G_{\mu\nu}\tilde{z}^{(0)\nu},
\nonumber\\
\tilde{U}_\mu^{(0)}&=&-2b_{\mu\nu}\tilde{z}^{(0)\nu}+G_{\mu\nu}z^{(0)\nu},
\end{eqnarray}
and therefore
\begin{equation}\label{eq:uexp}
V^{(0)\mu}(U^{(0)})=(g^{-1})^{\mu\nu}[2b_\nu^{\ \rho}U^{(0)}_\rho+\tilde{U}^{(0)}_\nu]=z^{(0)\mu},
\end{equation}
and consequently $\beta^{\pm}_{\mu}[V^{(0)}(U)]=\beta^{\pm}_{\mu}[z^{(0)}]$.
Substituting (\ref{eq:uexp}) into (\ref{eq:upmz}),
we obtain its solution
\begin{equation}\label{eq:usolwc}
u_{\pm\mu}=
-2\Pi_{\mp\mu\nu}[\Delta z^{(0)}]\partial_{\pm}z^\nu\mp
2\beta^{\mp}_{\mu}[z^{(0)}],
\end{equation}
with $\Delta z^{(0)\mu}=z^{(0)\mu}(\xi)-z^{(0)\mu}(\xi_{0}).$
Comparing it with (\ref{eq:usol}),
we find the
T-duality transformation law of the variables
\begin{equation}\label{eq:ct}
\partial_{\pm}y_\mu\cong
-2\Pi_{\mp\mu\nu}[\Delta z^{(0)}]\partial_{\pm}z^\nu\mp
2\beta^{\mp}_{\mu}[z^{(0)}].
\end{equation}
Note that this is the inverse transformation of (\ref{eq:tl}).
More precisely,
substituting $\partial_\pm y_\mu$ from (\ref{eq:ct}) into (\ref{eq:tl})
and using (\ref{eq:uexp}),
which is the consequence of the zeroth order of (\ref{eq:ct}),
one obtains
$\partial_{\pm}x^\mu=\partial_{\pm}z^\mu$.

Substituting (\ref{eq:usolwc}) into
the action (\ref{eq:afix}),
we obtain the action
\begin{equation}
^\star S_{fix}[z]=\kappa\int d^{2}\xi
\partial_{+}z^\mu \Pi_{+\mu\nu}[z(\xi)-z(\xi_{0})] \partial_{-}z^\nu,
\end{equation}
which is as we have learned in the section \ref{sec:construction}
invariant under the global shift
in the coordinate.
So, we can omit the term $z(\xi_{0})$ and obtain
the T-dual of the T-dual action
\begin{equation}\label{eq:ttdual}
^{\star\star}S[z]\equiv\,
^\star S_{fix}[z]=\kappa\int d^{2}\xi
\partial_{+}z^\mu \Pi_{+\mu\nu}[z] \partial_{-}z^\nu,
\end{equation}
which is in fact the initial action.
So, the second T-duality turns the doubled
target space $(y_\mu,\tilde{y}_\mu)$
back to the conventional space $z^\mu$.

Similarly as in the \ref{sec:dodatak},
it can be shown that the equation of motion of the T-dual of the T-dual action (original action) (\ref{eq:ttdual})
is the same as the
equation of motion of the gauge fixed action (\ref{eq:afix})
after elimination of the gauge fields on their equations of motion
\begin{equation}
\partial_{+}\partial_{-}z^\mu-B^\mu_{\ \nu\rho}\partial_{+}z^\nu\partial_{-}z^\rho=0.
\end{equation}

%%%%%%%%%%%%%%%%%%%%%%%%%%%%%%%%%%%%%%%%%%%%%%%%%%%%%%%%%%%%%%%%%%%%%%%%%%%%%%%%%%%
\section{The features of the T-duality}
\cleq

There are two important features of the T-duality which we
will consider here.
First, the momentum and the winding numbers
of the original theory are equal to the winding and
the momentum numbers of the T-dual theory
respectively.
Second,
the equation of motion and the Bianchi identity
of the original theory are equal to the Bianchi identity and
the equation of motion of the T-dual theory \cite{ALLP,Duff,GR}.
So, T-duality interchanges momentum and winding numbers,
as well as the equations of motion and Bianchi identities.
Because in our case,
the action is invariant up to the total divergences,
the conserved charges can in general differ
from the corresponding momenta.

%%%%%%%%%%%%%%%%%%%%%%%%%%%%%%%%%%%%%%%%%%%%%%%%%%%%%%%%%%%%%%%%%%%%%%%%%%%%%%%%%%%%%%%

\subsection{T-dualities in terms of the conserved currents and charges}

We will discuss the above mentioned features
by investigating the Noether and the topological
currents and their charges.
As a consequence of the global shift invariance
of the action (\ref{eq:action1})
there exist the conserved Noether currents
\begin{equation}
\partial_\alpha j^{\alpha}_{\mu}=0,
\end{equation}
 of the form
\begin{equation}\label{eq:netstr}
j^{\alpha}_{\mu}=\kappa\Big{[}
\Big{(}\eta^{\alpha\beta} G_{\mu\nu}+2\epsilon^{\alpha\beta}B_{\mu\nu}[x]
\Big{)}
\partial_\beta x^\nu-\beta^{\alpha}_{\mu}[x]
\Big{]},
\end{equation}
where
$\beta^{\alpha}_{\mu}$ is defined in (\ref{eq:etabeta}).
In the light-cone coordinates they have the following form
$j^{\pm}_{\mu}=\frac{1}{2}(j^{0}_{\mu}\pm j^{1}_{\mu})$
\begin{equation}
j^{\pm}_{\mu}=
\pm\kappa\Pi_{\pm\mu\nu}[x]
\partial_{\mp}x^\nu
-\kappa\beta^{\pm}_{\mu}[x].
\end{equation}
The current conservation equation $\partial_{+}j^{+}_{\mu}+\partial_{-}j^{-}_{\mu}=0$
is in fact the equation of motion of the original theory
\begin{equation}
\partial_{+}\partial_{-}x^\mu-B^{\mu}_{\ \nu\rho}\partial_{+}x^\nu\partial_{-}x^\rho=0.
\end{equation}

Let us now turn to the $T$-dual description.
As the consequence of (\ref{eq:ct}) one has
\begin{equation}\label{eq:jtran}
j^{\alpha}_{\mu}
\cong\,
{^\star i^{\alpha}_{\mu}}=-\kappa\epsilon^{\alpha\beta}
\partial_{\beta}y_\mu,
\end{equation}
where ${^\star i^{\alpha}_{\mu}}$ is the topological current, because
\begin{equation}\label{eq:icon}
\partial_\alpha\,{^\star i^{\alpha}_{\mu}}=0
\end{equation}
is the identity, which
is known as the Bianchi identity.
So, $T$-duality relates
the conservation of the Noether and the topological currents laws,
which are in fact
the equations of motion and the Bianchi identities.

Note that from (\ref{eq:netstr}) and (\ref{eq:jtran}),
for $\alpha=0$ one has
\begin{equation}\label{eq:pspb}
\pi_\mu-\kappa\beta^{0}_{\mu}[x]\cong
\kappa y^{\prime}_{\mu},
\end{equation}
where $\pi_\mu$ is canonical momentum
corresponding to the variable $x^\mu$
\begin{equation}
\pi_\mu=
\kappa(G_{\mu\nu}{\dot{x}}^\nu-2B_{\mu\nu}[x]x^{\prime\nu}).
\end{equation}

The charges associated with the conserved currents (\ref{eq:netstr})
\begin{equation}
Q_\mu=\int_{-\pi}^{\pi}d\sigma j^{0}_{\mu}
=\int_{-\pi}^{\pi}d\sigma\Big{[} \pi_\mu
-\kappa\beta^{0}_{\mu}\Big{]},
\end{equation}
in general could differ from
the momenta quantum numbers (Kaluza-Klein modes).
The difference is the infinitesimal part of the momenta.
Its mode expansion corresponds to the expressions
(3.21) and (3.22) of Ref. \cite{ALLP}.
In the particular case,
when the string is curled up around only one compactified dimension,
i.e. $x^{i}=c\sigma,\,x^{j}=0,\,j\neq i$
one has $\beta^{0}_{\mu}=0$,
because of the antisymmetry of $B_{\mu\nu\rho}$.
Then the conserved charges turn to momenta quantum numbers.

The charge corresponding to the conserved topological current
\begin{equation}
^\star q_\mu=\int_{-\pi}^{\pi}d\sigma\,
{^\star i^{0}_{\mu}}=\kappa
\int_{-\pi}^{\pi}d\sigma y^{\prime}_{\mu},
\end{equation}
is just the winding number of the $T$-dual theory. As the consequence of (\ref{eq:pspb}),
the Noether charges transform under $T$-duality into the topological charges
\begin{equation}\label{eq:qq}
Q_\mu\cong {^\star q}_\mu.
\end{equation}
For $\beta^{0}_{\mu}=0$ this just describes
the fact that $T$-duality
transforms the momenta numbers of the initial theory
into the winding numbers of the $T$-dual theory.

Because the T-dual of the T-dual theory is the original theory,
we can apply the same procedure in the other direction.
From (\ref{eq:dualna}) we obtain
the T-dual Noether currents
\begin{eqnarray}\label{eq:jzvezda}
{^\star j^{\alpha\mu}}&=&
\kappa
\Big{[}
\Big{(}
\eta^{\alpha\beta}(G^{-1}_{E})^{\mu\nu}
[\Delta V]
+\kappa\epsilon^{\alpha\beta}\theta^{\mu\nu}
[\Delta V]
\Big{)}\partial_\beta y_\nu
-(g^{-1})^{\mu\nu}\epsilon^{\alpha}_{\ \beta}\beta^{\beta}_{\nu}[V]
-\kappa\theta_{0}^{\mu\nu}\beta^{\alpha}_{\nu}[V]\Big{]}.
\nonumber\\
\end{eqnarray}
In the light-cone coordinates one has
\begin{equation}\label{eq:jzvezdad}
{^\star j^{\pm\mu}}=\pm\frac{\kappa^{2}}{2}\Theta^{\mu\nu}_{\mp}[\Delta V]
\Big{[}\partial_{\mp}y_\nu\mp2\beta^{\pm}_{\nu}
[V]\Big{]}.
\end{equation}
The conservation law for T-dual current
\begin{equation}
\partial_{+}\,{^\star j^{+\mu}}+
\partial_{-}\,{^\star j^{-\mu}}=0,
\end{equation}
is just the equation of motion in
T-dual theory
\begin{eqnarray}
\partial_{+}\Big{[}
\Theta^{\mu\nu}_{-}[\Delta V]\partial_{-}y_\nu
-2\Theta^{\mu\nu}_{0-}\beta^{+}_\nu[V]
\Big{]}
-\partial_{-}\Big{[}
\Theta^{\mu\nu}_{+}[\Delta V]\partial_{+}y_\nu
+2\Theta^{\mu\nu}_{0+}\beta^{-}_\nu[V]
\Big{]}=0.
\end{eqnarray}

T-duality according to (\ref{eq:tl})
transform Noether currents of the T-dual theory to the
topological currents of the original theory
(formally to the topological currents of T-dual of T-dual theory)
\begin{equation}\label{eq:tnt}
{^\star j^{\alpha\mu}}\cong
i^{\alpha\mu}=-\kappa\epsilon^{\alpha\beta}
\partial_\beta x^\mu.
\end{equation}
The conservation of the topological currents $\partial_\alpha i^{\alpha\mu}=0$
are just the Bianchi identities.
From (\ref{eq:jzvezda}) and (\ref{eq:tnt}) for $\alpha=0$ follows
\begin{equation}\label{eq:pspbd}
{^\star \pi^{\mu}}
-\kappa^{2}\theta_{0}^{\mu\nu}\beta^{0}_{\nu}[V]
\cong\kappa x^{\prime\mu},
\end{equation}
where ${^\star \pi^{\mu}}$ is canonical momentum in the T-dual theory
\begin{eqnarray}
^\star\pi^\mu
=
\kappa(G^{-1}_{E})^{\mu\nu}\big[\Delta  V[y]\big]
\dot{y}_\nu
-\kappa^{2}\theta^{\mu\nu}\big[\Delta  V[y]\big]
y^\prime_\nu
-\kappa  (g^{-1})^{\mu\nu}\beta^1_\nu\big[V[y]\big].
\end{eqnarray}

The conserved charges of the dual Noether and the original topological currents
for $\beta^{\alpha}_{\mu}=0$
\begin{eqnarray}
&&
{^\star Q^\mu}=
\int_{-\pi}^{\pi}d\sigma\,{^\star j^{0\mu}}
=\int_{-\pi}^{\pi}d\sigma\,{^\star \pi^\mu},
\nonumber\\
&&
{q^\mu}=
\int_{-\pi}^{\pi}d\sigma\,{i^{0\mu}}
=\kappa\int_{-\pi}^{\pi}d\sigma\, x^{\prime\mu},
\end{eqnarray}
are momenta modes of the dual theory and the winding modes
of the original theory. They are also
according to (\ref{eq:pspbd}),
connected by the T-duality transformation
\begin{equation}
{^\star} Q^\mu\cong q^\mu.
\end{equation}

In the following table we summarize the obtained relations:
T-duality transformation relates the Noether currents
with the topological ones;
the corresponding conservation laws relate
the equations of motion
with Bianchi identities,
while the corresponding Noether cha\-rg\-es relate momenta and winding modes.\\

\begin{tabular}{|l| l| l|}
\hline
{\it Original theory} $\quad S$ &$\longrightarrow$  & {\it T-dual theory} $\quad ^\star S$\\
\hline
{\it Noether current} $\quad j^{\alpha}_{\mu}\quad$
&& {\it Topological current} $\quad{^\star i^{\alpha}_{\mu}}
=-\kappa\epsilon^{\alpha\beta}\partial_\beta y_\mu$
\\
{\it Conservation law = Equation of motion}
 & & {\it Conservation law = Bianchi identity}
\\
$\quad \partial_\alpha j^{\alpha}_{\mu}=0$ & &
$\quad\partial_\alpha{^\star i^{\alpha}_{\mu}}=0$
\\
{\it Noether conserved charge}  && {\it Topological conserved charge}\\
$Q_\mu=\int_{-\pi}^{\pi}d\sigma j^{0}_{\mu}
=\int_{-\pi}^{\pi}d\sigma \pi_\mu=P_\mu$
&&$^\star q_\mu=\int_{-\pi}^{\pi}d\sigma\,
{^\star i^{0}_{\mu}}=\kappa
\int_{-\pi}^{\pi}d\sigma y^{\prime}_{\mu}=\,^\star W_\mu$\\
\hline
{\it T-dual of T-dual theory} $\quad^{\star\star}S=S$ &$\longleftarrow$& {\it T-dual theory} $\quad ^\star S$\\
\hline
%%%%%%%%%%%%%%%%%%%%%%%%%%%%%%%%%%%%%%%%%%%%%%%%%%%%%%%%%%%%%%%%%%%%%%%%%%%%%%%%%%%%%%%%%%%%%%%%%%%%%%%%%
{\it Topological current}
$\quad i^{\alpha\mu}=-\kappa\epsilon^{\alpha\beta}\partial_\beta x^\mu$
&& {\it Noether current}$\quad{^\star j^{\alpha\mu}}$\\
{\it Conservation law = Bianchi identity}
 &&
{\it Conservation law = Equation of motion}
\\
$\quad\partial_\alpha i^{\alpha\mu}=0$ &&
$\quad\partial_\alpha{^\star j^{\alpha\mu}}=0$
\\
{\it Topological conserved charge} && {\it Noether conserved charge}\\
$q^\mu=\int_{-\pi}^{\pi}d\sigma\,
{ i^{0\mu}}=\kappa
\int_{-\pi}^{\pi}d\sigma x^{\prime\mu}=W^\mu$&&
$^\star{Q}^\mu=\int_{-\pi}^{\pi}d\sigma\,{^\star j^{0\mu}}
=\int_{-\pi}^{\pi}d\sigma\,{^\star\pi}^\mu=\,^\star P^\mu$
\\
\hline
\end{tabular}

%%%%%%%%%%%%%%%%%%%%%%%%%%%%%%%%%%%%%%%%%%%%%%%%%%%%%%%%%%%%%%%%%%%%%%%%%%%%%%%%%%%%%%%
\section{Conclusion}
\cleq

In this paper,
we considered the closed bosonic string moving in the
weakly curved background.
This background is defined by the constant
space-time metric and the linear in coordinate
Kalb-Ramond field,
where the coordinate dependence is infinitesimally small.
With such a choice the space-time equations of motion
were satisfied.
The aim of the paper was to investigate the T-dual theory
in curved background.

Earlier, in number of papers the similar topic,
restricted to the string in the flat background, was discussed.
In these papers,
the prescriptions for the construction
of the T-dual theories were established.
Here we presented the generalization of the covariant 
Bu\-scher's construction.

In Buscher's construction, one starts with the sigma model
constructed from background fields $G,B,\Phi$
which do not
depend on some coordinates $x^{a}$. So, the corresponding abelian
isometries leave the action invariant. We started with the sigma
model in the weakly curved background. We found that the action
still has the global symmetry $\delta x^\mu=\lambda^\mu=const$,
even-though the background fields depend of these coordinates. 
So,
we gauged it in the usual way by introducing the gauge fields
$v^\mu_\alpha$, replacing the derivatives $\partial_{\alpha}x^\mu$
with the covariant ones.
In our case this was not sufficient to construct the invariant
action, because the background field $B_{\mu\nu}$ depends on
$x^\mu$ which is not gauge invariant. 
The essential new step in our gauging prescription is the introduction of
the invariant coordinate,
as the line integral (primitive) of its covariant derivatives.
This kind of coordinate enables the local invariance.
It remains to find out,
to which class of backgrounds this generalized procedure
can be applied.

As usual, for the T-dual theory to be physically equivalent
to the original theory, one had to eliminate all the degrees of
freedom carried by the gauge fields. This was achieved by adding
the Lagrange multiplier term $y_{\mu}F^{\mu}_{01}$ into
Lagrangian. At this point we fixed the gauge. The action obtained
in this way reduced to the initial one
on the equations of motion for the
Lagrange multiplier $y_\mu$.

The Lagrange multiplier term $y_{\mu}F^{\mu}_{01}$
guarantees that the gauge field is closed ($dv=0$)
but one should consider the topological contribution as well.
Because of this, the additional investigation of the holonomies of $v$,
should be performed.
In order to solve these problems,
connected with the global structure of the theory,
 following Refs. \cite{RV,AABL,RP,GR,QT}
we will consider the quantum theory
in some of  the further papers.

On the solution of equations of motion for the gauge fields,
one obtains the T-dual action.
In the flat background case,
the T-dual action was given in terms of the T-dual variable
which turned out to be the Lagrange multiplier itself.
In the weakly curved case,
the T-dual action is defined in the doubled space given in terms
of the Lagrange multiplier and its $T$-dual in the flat space.
The dual background fields depend on $\Delta V^\mu$, the linear combination
of these variables.

Starting from the T-dual action and
following the T-dual prescription we proposed,
we obtained the initial one.
The fact that we succeeded in finding the rules for
obtaining the T-dual theory from the initial one in the weakly
curved background and the inverse rules,
allowed us to comment
the two main features of T-duality.
These two are:
the T-duality relates the original and the T-dual theory,
by mapping
the momentum numbers of one theory with the winding numbers
of the other and
the equations of motion of one with the Bianchi identities
of the other.

Starting with the initial theory and its global shift invariance
we found the conserved Noether currents $j^\alpha_\mu$.
The currents conservation laws
$\partial_\alpha j^\alpha_\mu=0$
are in fact the equations of motion
and the associated charges
$Q_\mu=\int_{-\pi}^{\pi}d\sigma j^{0}_{\mu}=P_\mu$
are momentum numbers.
T-dual of Noether currents are the topological currents $^\star i^\alpha_\mu$ of T-dual theory.
Their conservation laws $\partial_\alpha\, ^\star i^\alpha_\mu=0$
are in fact Bianchi identities and the associated charges
$^\star q_\mu=\int_{-\pi}^{\pi}\,^\star i^{0}_\mu
=\,{^\star W_\mu}$
are the winding numbers.
So, T-duality relates equations of motion with Bianchi identities
and the momentum with winding modes.
Analogously,
starting with the T-dual action and its global symmetry
we found that
the Noether currents and their associated charges
(dual momentum numbers)
correspond to the topological currents of the initial theory
and theirs charges (winding numbers).
%%%%%%%%%%%%%%%%%%%%%%%%%%%%%%%%%%%%%%%%%%%%%%%
\appendix
%%%%%%%%%%%%%%%%%%%%%%%%%%%%%%%%%%%%%%%%%%%%%%%
\section{T-dual equation of motion}\label{sec:dodatak}

Let us demonstrate that the T-dual equation of motion can be obtained
from both T-dual action (\ref{eq:dualna}) and the gauge fixed action (\ref{eq:sfix}).
The T-dual action depends on $y_\mu$,
so there is one equation of motion which depends on $y_\mu$.
On the other hand,
the gauge fixed action
beside $y_\mu$ depends on the gauge fields $v^\mu_\pm$ as well.
Treating $v^\mu_\pm$
as auxiliary fields,
we will use their equations of motion to eliminate them from the third equation
and obtain one $y_\mu$ dependent equation.

\subsection{The equation of motion for the gauge fixed action} 

Let us find the equations of motion for the gauge fixed action
(\ref{eq:sfix}).
This will be done iteratively.
Equations of motion, obtained varying by the finite parts $(v_{\pm}^{(0)\mu},y_\mu^{(0)})$
introduced in (\ref{eq:nulajedan}),
contain both finite and infinitesimal part.
The finite part of these equations is
\begin{subequations}
\begin{eqnarray}\label{eq:zero}
&&\Pi_{0+\mu\nu}v^{(0)\nu}_{-}+\frac{1}{2}\partial_{-}y^{(0)}_\mu=0,
\\\label{eq:zero1}
&&\Pi_{0-\mu\nu}v^{(0)\nu}_{+}+\frac{1}{2}\partial_{+}y^{(0)}_\mu=0,
\\\label{eq:zero2}
&&\partial_{+}v^{(0)\mu}_{-}
-\partial_{-}v^{(0)\mu}_{+}=0.
\end{eqnarray}
\end{subequations}
The equations of motion obtained varying by the infinitesimal parts $(v_{\pm}^{(1)\mu},y_\mu^{(1)})$
are identical to (\ref{eq:zero}), (\ref{eq:zero1}), (\ref{eq:zero2}).
The equation
(\ref{eq:zero2}) guarantees that (\ref{eq:vdef})
in the zeroth order
does not depend on the choice of the path $P$.
So, we can write
\begin{equation}\label{eq:vdefxi}
\Delta V^{(0)\mu}=V^{(0)\mu}(\xi)-V^{(0)\mu}(\xi_{0}),
\end{equation} 
and 
\begin{equation}\label{eq:vV}
\partial_\alpha V^{(0)\mu}=
v^{(0)\mu}_\alpha.
\end{equation}

The solution of (\ref{eq:zero}) and (\ref{eq:zero1}) is
\begin{equation}
v_{\pm}^{(0)\mu}(y)=
-\kappa\,{\Theta}^{\mu\nu}_{0\pm}
\partial_{\pm}y_\nu^{(0)},
\end{equation}
where ${\Theta}^{\mu\nu}_{0\pm}$ is defined in (\ref{eq:tvnula}).
Using the zeroth order value of $V^\mu$
we can rewrite (\ref{eq:v0}) as 
\begin{eqnarray}
&&v_{\pm}^{(0)\mu}(y)=\partial_\pm
V^{(0)\mu}(y),
\nonumber\\
&&V^{(0)\mu}=
-\kappa\,{\theta}^{\mu\nu}_{0} y_\nu^{(0)}
+(g^{-1})^{\mu\nu}\tilde{y}_\nu^{(0)}.
\end{eqnarray}

Let us now turn to the infinitesimal part of the
equations of motion obtained varying $S_{fix}$ by the finite parts $(v_{\pm}^{(0)\mu},y_\mu^{(0)})$.
They are equal to
\begin{subequations}
\begin{eqnarray}\label{eq:first}
&&h_{\mu\nu}[\Delta V^{(0)}]v^{(0)\nu}_{-}
+\Pi_{0+\mu\nu}v_{-}^{(1)\nu}
+\frac{1}{2}\partial_{-}y^{(1)}_\mu=
\beta^{+}_\mu[V^{(0)}],
\\\label{eq:first1}
&&h_{\mu\nu}[\Delta V^{(0)}]v^{(0)\nu}_{+}
+\Pi_{0-\mu\nu}v_{+}^{(1)\nu}
+\frac{1}{2}\partial_{+}y^{(1)}_\mu=
-\beta^{-}_\mu[V^{(0)}],
\\\label{eq:first2}
&&\partial_{+}v^{(1)\mu}_{-}
-\partial_{-}v^{(1)\mu}_{+}=0.
\end{eqnarray}
\end{subequations}
Solving the first two equations,
with the help of (\ref{eq:v0}) and (\ref{eq:ptnula}),
we get
\begin{eqnarray}\label{eq:v1}
v^{(1)\mu}_\pm&=&
-\kappa\Theta_{0\pm}^{\mu\nu}\partial_\pm y^{(1)}_\nu
-\kappa\Theta_{1\pm}^{\mu\nu}[\Delta V^{(0)}]\partial_\pm y^{(0)}_\nu
\mp
2\kappa\Theta_{0\pm}^{\mu\nu}\beta_\nu^{\mp}[V^{(0)}],
\end{eqnarray}
where
\begin{equation}\label{eq:theta1}
\Theta_{1\pm}^{\mu\nu}[x]=
-2\kappa \Theta_{0\pm}^{\mu\rho}h_{\rho\sigma}[x]\Theta_{0\pm}^{\sigma\nu}.
\end{equation}
So, the solution for $v_\pm^\mu$ is just the equation (\ref{eq:vsol1})
\begin{eqnarray}\label{eq:vpmres}
v_{\pm}^{\mu}[y]&=&
-\kappa\,{\Theta}^{\mu\nu}_{\pm}[\Delta V^{(0)}]
\partial_{\pm}y_\nu
\mp2\kappa {\Theta}^{\mu\nu}_{0\pm}\beta^{\mp}_\nu[V^{(0)}],
\end{eqnarray}
where ${\Theta}^{\mu\nu}_{\pm}$,
defined in (\ref{eq:velikateta}),
is the inverse of 
$\Pi_{\pm\mu\nu}$ (see the relation
(\ref{eq:tp})).

Finally,
(\ref{eq:zero2}) and (\ref{eq:first2})
produce
$\partial_{+}v^{\mu}_{-}
-\partial_{-}v^{\mu}_{+}=0$.
On the solution (\ref{eq:vpmres}) this equation becomes
\begin{eqnarray}\label{eq:eqyv}
&&
\partial_{+}\Big{[}
{\Theta}^{\mu\nu}_{-}[\Delta V^{(0)}]
\partial_{-}y_\nu
-2{\Theta}^{\mu\nu}_{0-}\beta^{+}_\nu[V^{(0)}]
\Big{]}
-\partial_{-}\Big{[}
{\Theta}^{\mu\nu}_{+}[\Delta V^{(0)}]
\partial_{+}y_\nu
+2{\Theta}^{\mu\nu}_{0+}\beta^{-}_\nu[V^{(0)}]
\Big{]}=0.
\nonumber\\
\end{eqnarray}
This is the equation of motion of the T-dual theory.

%%%%%%%%%%%%%%%%%%%%%%%%%%%%%%%%%%%%%%%%%%%%%%
\subsection{The equation of motion for the T-dual action}

Again, we will find the equations of motion iteratively.
Variation
of the T-dual action (\ref{eq:dualna}),
 with respect to $y_\mu$ produces in the zeroth order
\begin{equation}
\partial_{+}
\partial_{-}y_\mu^{(0)}=0.
\end{equation}
On this equation $\Delta \tilde{y}_\mu$ becomes path independent and we have
$
\partial_\pm V^{(0)\mu}=-\kappa\Theta_{0\pm}^{\mu\nu}\partial_\pm y_\nu
$.
Variation by $V^\mu$,
with the help of (\ref{eq:theta1}), gives
\begin{eqnarray}
\delta_{V}{^\star S}
=
\kappa^{2} \int d\xi^{2}
\Big{[}
\Theta_{0-}^{\mu\nu}
\partial_{+}\beta^{+}_\nu[V^{(0)}]
+\Theta_{0+}^{\mu\nu}\partial_{-}\beta^{-}_\nu[V^{(0)}]
\Big{]}\delta y_\mu,
\end{eqnarray}
where $\beta^\pm_\mu$ are defined in (\ref{eq:betapm})
and so the equation of motion for $y_\mu$ is 
indeed the eq. (\ref{eq:eqyv}).

%%%%%%%%%%%%%%%%%%%%%%%%%%%%%%%%%%%%%%%%%%%%%%%%%

%%%%%%%%%%%%%%%%%%%%%%%%%%%%%%%%%%%%%%%%%%%%%%%%%%%%%%%%%%%%%%%%%%%
%%%%%%%%%%%%%%%%%%%%%%%%%%%%%%%%%%%%%%%%%%%%%%%%%%%%%%%%%%%%%%%%%%%
\end{document}